\documentclass[jkps,preprint,fleqn,showpacs,showkeys]{revtex4}
\usepackage{graphicx}
\usepackage{amssymb}
\usepackage{amsmath}
\usepackage{bm}
\begin{document}
\setcounter{page}{0}
\title[]{Substitution -- and Strain -- Induced Magnetic Phase Transition in Iron Carbide}
\author{D. Odkhuu}
\affiliation{Department of Physics, Incheon National University, Incheon 406-772, Republic of Korea}
\author{N. Tsogbadrakh}
\email{Tsogbadrakh@num.edu.mn}
\author{A. Dulmaa}
\author{N. Otgonzul}
\affiliation{Department of Physics, National University of Mongolia, Ulaanbaatar - 14201, Mongolia}
\author{D. Naranchimeg}
\affiliation{Department of Physics, Mongolian University of Science and Technology, Ulaanbaatar - 14191, Mongolia}
\date[]{Received 10 May 2016, in final form 30 September 2016}

\begin{abstract}
Cementite -- type carbides are of interest for magnetocaloric applications owing to temperature -- or pressure -- induced magnetic phase transition. Here, using first -- principles calculations, we investigate the magnetism and the magnetic phase transition in iron carbide (Fe$_3$C) with the substitution of Cr atoms at Fe sites with the strain effect.
The presence of Cr atoms is found to give rise to a second -- order magnetic phase transition from a ferromagnetic phase for Fe$_3$C to a nonmagnetic phase in chromium carbide (Cr$_3$C). While the ternary Fe$_2$CrC and Cr$_2$FeC compounds prefer the ferrimagnetic ground state, the magnitudes of both the Fe and the Cr spin moments, which are antiparallel in orientation,  decrease as x increases in Fe$_{3-x}$Cr$_x$C (x = 0, 1, 2, and 3). Furthermore, the fixed spin -- moment calculations indicate that the magnetization of Fe$_{3-x}$Cr$_x$C compounds can be delicately altered via the strain effect and that the magnetic -- nonmagnetic phase transition occurs at an early stage of Cr substitution, x = 2.
\end{abstract}



\maketitle

\begin{center}
\section{INTRODUCTION}
\end{center}

Pure iron and carbon containing iron are important metals of modern industries and technologies and exhibit rich phase diagrams. In particular, the engineering of the magnetic phase transition temperature in iron is of great interest in fundamental research and practical applications. Amongst the several ways to modulate the transition temperature, the doping or alloying with a semiconductor element is the most common and practical in iron and its compounds. Pure Fe is known to undergo a magnetic phase transition from a low temperature ferromagnetic (FM) to a high temperature paramagnetic (PM) phase at around 1043 K \cite{Spaldin}. This transition temperature can be further greatly reduced by the addition of carbon impurities. More specifically, the Curie temperature is decreased to 480 K at the lowest carbon content of 25\% C atomic weight in Fe, i.e., Fe$_3$C (called cementite), among all known Fe -- C compounds \cite{Smith1}. In a Fe K-edge X-ray magnetic circular dichroism (XMCD) experiment \cite{Duman}, the pressure -- induced magnetic phase transition from a FM  to a nonmagnetic (NM) phase was demonstrated in Fe$_3$C at room temperature. Furthermore, this iron carbide is one of the most important phases in steels and is highly desirable for soft magnet cores in transformer core steels \cite{Umemoto}, even though it is a hard and brittle material \cite{Smith2}.  During last few decade, the elastic and mechanical properties of Fe$_3$C have been intensively investigated in order to improve its ductility \cite{Shein, Wang, Lv, Zhang}. For instance, the substitution of Cr at Fe sites was explored for Fe$_2$CrC and Cr$_2$FeC compounds in a recent first -- principles study, and these ternary compounds were found to exhibit mixed  metallicity, covalency and ionicity behaviors \cite{Lv}.

On the other hand, a magnetic material that exhibits room -- temperature high magnetic entropy and magnetic phase transition associated with the magnetocaloric effect is highly indispensible for magnetic refrigeration and thermomagnetic power generation applications \cite{Wada, Tegus2}. Within this concept, the room -- temperature magnetic phase transition of Fe$_3$C with Cr substitution was explored in recent magnetic measurements \cite{Zhao}, and a second -- order phase transition was shown to be feasible at room -- temperature. Moreover, a number of experimental and theoretical studies have been done on the structural, mechanical, electronic, and magnetic properties of pristine cementite \cite{Shein, Wang, Lv, Zhang}, 
including with substitutions of Co and Ni atoms at Fe sites \cite{Shein}. Nevertheless, exploration of the influences of Cr substitution and the strain effect, as lattice mismatch occurs quite often during growth, on the magnetism and on the magnetic phase transition in Fe$_3$C is still missing.

In this study, by fixed spin -- moment (FSM) calculation, we present results for the electronic structure and the magnetic phase transition with and without the strain effect in Fe$_{3-x}$Cr$_x$C compounds, where x = 0, 1, 2, and 3 \cite{Williams, Schwarz, Moruzzi, Deniszczyk}. This computational technique allows the ground state of a constrained system \cite{Dederichs} to be determined in the plane -- wave self -- consistent method within the framework of density functional theory (DFT)  \cite{Hohenberg, Kohn}. Iron (x = 0) and chromium (x = 3) carbides have FM and NM ground -- state magnetic phases, respectively. At x = 1 and 2, the substitutional Cr atoms induce magnetic moments antiparallel to those of Fe,
for which the ferrimagnetic (FIM) configurations are energetically more favorable than the FM and the NM phases.
We also investigate the sensitivity of the magnetism and the magnetic ground -- state to the strain effect and found that the magnetic phase transition and the temperature can be modified via control of the epitaxial strain.

\qquad
\begin{center}
{\bf II. COMPUTATIONAL DETAILS}
\end{center}

\begin{figure}
\includegraphics[width=8.5cm]{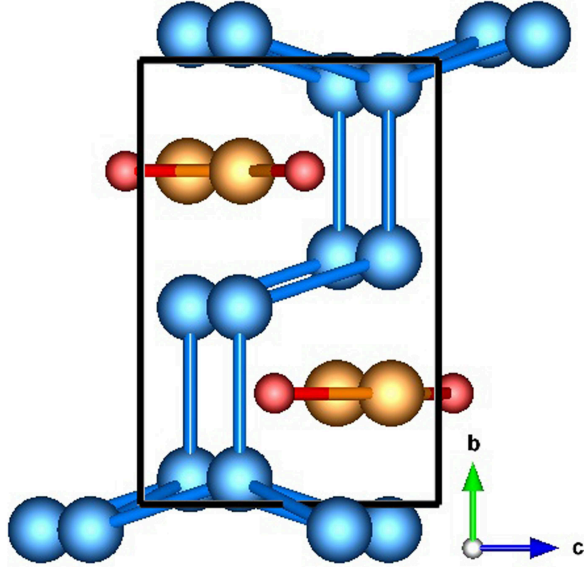}
\caption{(Color online) Side view of the unit cell for orthorhombic iron carbide (Fe$_3$C). The atomic species are denoted in spheres with different colors: larger orange and blue and smaller red spheres are Fe$_1$, Fe$_2$, and C atoms, respectively.}
\label{fig1}
\end{figure}

The iron carbide consists of four formula units (f.u.) in an orthorhombic unit cell, and the space group is Pnma (no. 62 in the International Tables for Crystallography) \cite{Fasiska, Wood}. The unit cell has four iron atoms, denoted as Fe$_1$, in the Wyckoff position 4c, eight iron atoms, denoted as Fe$_2$, in the Wyckoff position 8d, and four carbon atoms in the interstices, as shown in Figure \ref{fig1}. The same orthorhombic unit cell is used for all Fe$_{3-x}$Cr$_x$C systems considered in the present study, because both the iron and the chromium carbides have the same structure \cite{Medvedeva}.
Our calculations are based on the plane -- wave self -- consistent field (PWscf) method using the generalized gradient approximation (GGA) by Perdew, Burke and Ernzerhof (PBE) \cite{Perdew} within the framework of DFT \cite{Hohenberg, Kohn}, as implemented in the QUANTUM ESPRESSO package \cite{Gianmozzi}. The interaction between the ions and valence electrons is expressed as the ultrasoft pseudopotential \cite{Vanderbilt}. The following electronic states are treated as valence states: C(2s$^2$, 2p$^2$), Cr(3s$^2$, 3p$^6$, 3d$^5$, 4s$^1$) and Fe(3s$^2$, 3p$^6$, 3d$^6$, 4s$^2$). The wave functions are expressed as plane waves up to a kinetic energy cutoff of 30 Ry. The summation of charge densities is carried out using the special \emph{k} -- points generated by the (7 x 5 x 6) grids of the Monkhorst -- Pack scheme \cite{Monkhorst}. The tetrahedral method \cite{Blochl} is used when the electronic density of state (DOS) is evaluated. In order to obtain optimized atomic structures, we allow the ionic positions and the lattice parameters to be fully relaxed until the residual forces are less than 0.05 eV/$\textrm{\AA}$ for each atom. In the strain calculation, both tensile and compressive strains up to $\pm$ 5$\%$ along the short coordinate z -- axis are taken into account while the xy -- plane lattice is unchanged.
The occupation numbers of electrons are expressed as a Gaussian distribution function with an electronic temperature of $kT$ = 0.001 Ry. The FSM method \cite{Williams, Schwarz, Moruzzi, Deniszczyk} is used for the magnetic phase transition calculations, which yields the ground state of a constrained system \cite{Dederichs}. In the FSM calculation, the value of the parameter used for the constrained magnetization was chosen as 5 Ry \cite{Gianmozzi}. 
 
\begin{center}
{\bf III. RESULTS AND DISCUSSION}
\end{center}

Before investigating the magnetic properties, we first optimized the atomic structures of Fe$_{3-x}$Cr$_x$C for each x.
As shown in Table 1, the optimized lattice parameters \emph{a}, \emph{b}, and \emph{c} of the orthorhombic structure in Fe$_{3}$C (Cr$_{3}$C) carbide are 5.05 (5.18), 6.73 (6.65), and 4.49 (4.51) \AA, respectively. Those of the ternary Fe$_2$CrC (Cr$_2$FeC) compound are calculated to be 5.04 (5.13), 6.88 (6.57), and 4.40 (4.51) \AA, respectively. Overall, our calculated structural properties reproduce quite well the experimental \cite{Fasiska, Wood, Inoue} and previous
theoretical results \cite{Lv}, which are shown in parentheses in Table 1. Note that the structural parameters of cementite (Fe$_{3}$C) are not strongly altered in the presence of Cr substitution due to atomic radii of the Fe and Cr transition metals being similar. With the optimized lattice for each x in Fe$_{3-x}$Cr$_x$C, we then calculated the total energies of three different magnetic configurations, namely, the FM, FIM, and NM configurations. The magnetic energies, calculated as the energy difference between the FM/FIM and NM states ($\Delta$E = E$_{\textrm{tot}}$[FM/FIM] -- E$_{\textrm{tot}}$[NM]),
and the magnetic moments for each atomic site are shown in Table 1 for the most stable magnetic structure of FM for Fe$_3$C and of FIM for Fe$_2$CrC and Cr$_2$FeC. As seen in Table 1, the value of $\Delta$E decreases with increasing x and is eventually quenched at x = 3, which indicates the NM ground state of Cr$_3$C. The FM nature of Fe$_3$C is evidenced by the value of $\Delta$E being about -2.5 eV/cell. Similar results were also found in previous studies \cite{Shein}. The FM state of Fe$_3$C is more favorable than its FIM state because of the energy difference of --1.5 eV/cell. Two individual Fe sites, Fe$_1$ and Fe$_2$, have slightly different magnetic moments of about 2.09 and 2.05 $\mu_B$, respectively, which are smaller than that (2.2 $\mu_B$) in bulk body-centered cubic Fe. This reduced magnetic moment, as well as the small difference between the Fe$_1$ and the Fe$_2$ sites, should be attributed to the presence of carbon atoms, which will be addressed in the discussion on the electronic structure. Furthermore, C atoms have an induced moment of --0.24 $\mu_B$, antiparallel to the Fe moments. In Fe$_2$CrC, as the four Fe$_1$ ions in the Wyckoff position 4c are replaced by Cr atoms, the FIM structure is found to be stabilized in the self -- consistent total energy calculations, not the FM phase, and is more stable than the NM phase due to its having a magnetic energy of --0.80 eV/cell. In this FIM structure, the substitutional Cr atoms have an induced magnetic moment of --1.25 $\mu_B$ antiparallel to the Fe$_2$ moment (1.65 $\mu_B$). In the Cr$_2$FeC compound, we replace the eight Fe$_2$ ions in Wyckoff position 8d with Cr atoms, in which case the FIM phase is still found to be more favorable by about --0.13 eV/cell than the NM phase. In addition, the total energies of the FIM states for the Fe$_2$CrC and the Cr$_2$FeC compounds are degenerate with respect to their FM states while all magnetic states for Cr$_3$C are degenerate with respect to the ground state of the NM phase. The magnetic stability and energies for iron and chromium carbides are in agreement with the results obtained by Konyaeva and Medvedeva \cite{Konyaeva}. Moreover, Shein \textit{et al.} reported that the Fe$_3$C and the Co$_3$C compounds exhibited FM state, while the ground state of the Ni$_3$C compound was PM \cite{Shein}.

\begin{figure}
\includegraphics[angle=-90, width=10.0cm]{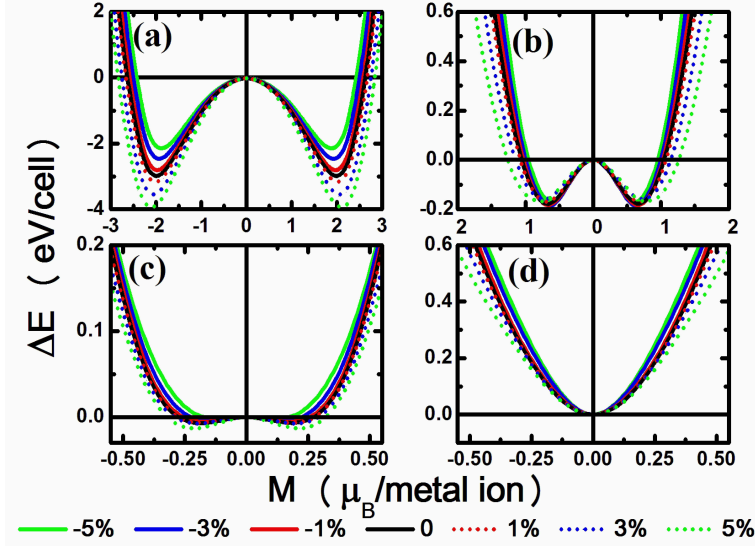}
\caption{(Color online) Magnetic energy $\Delta$E, defined as the total energy difference between the FM/FIM and the NM phases ($\Delta$E=E$_{\textrm{tot}}$[FM/FIM] -- E$_{\textrm{tot}}$[NM]), as a function of the magnetization M for (a) Fe$_3$C, (b) Fe$_{2}$CrC, (c) Cr$_{2}$FeC, and (d) Cr$_3$C with compressive (solid) and tensile (dotted) strains.}
\label{fig2}
\end{figure}

$\Delta$E as a function of the magnetization for Fe$_{3-x}$Cr$_x$C with x = 0, 1, 2, and 3 is plotted in Figure \ref{fig2}.
Figure \ref{fig2}(a) shows two local minima in energy at a finite magnetization of about $\pm$1.47 $\mu_B$ per atom and a saddle point at M = 0 in Fe$_3$C, indicating a FM ground state. As x is increased in Fe$_{3-x}$Cr$_x$C, these local minima shift towards zero net moment, and the absolute value of $\Delta$E decreases. At x = 3, the minima entirely overlap at M = 0, which ensures a NM ground state of Cr$_3$C, as shown in Table 1. From these results, one can argue that the Fe$_{3-x}$Cr$_x$C compounds undergo a second -- order phase transition as Fe$_3$C goes to Cr$_3$C \cite{Martin}.

Enhanced and induced magnetic properties in transition metals due to strain engineering are a subject of great study today  \cite{odkhuu1,odkhuu2}. For instance, we show in our previous studies that ferromagnetism can be achieved in metastable Ru metal under an epitaxial strain imposed by an underlying Mo(110) substrate \cite{odkhuu2}. Furthermore, a high magnetic entropy and a large thermal hysteresis, which are associated with a second -- order magnetic phase transition, have been found in MnFeP$_{1-x}$T$_{x}$ (T = P, As, Si, Ge ) alloys, where the coupling between the magnetization and the strain plays an important role \cite{Tegus1}. Therefore, we here explore the effect of strain on the magnetism, particularly the magnetic phase transition, in Fe$_{3-x}$Cr$_x$C compounds. In Fig. 2, $\Delta$E as a function of the magnetization in Fe$_{3-x}$Cr$_x$C compounds with x = 0, 1, 2, and 3 is shown for different compressive (dotted line) and tensile (solid line) strains. As shown in Figure \ref{fig2}(a), the magnetization per atom of Fe$_3$C decreases with incresing compressive strain whereas tensile strains enhance the magnetic moments, as well as the stability of the FM state. The $\Delta$E versus magnetization at x = 1 [Figure \ref{fig2}(b)] and x = 2 [Figure \ref{fig2}(c)] exhibit a trend similar to that in Fe$_3$C. Further, the magnetism of Cr$_2$FeC completely disappears when the compressive strain reaches -5\%, as shown in Figure \ref{fig2}(c). No significant changes in the stability of the magnetic phase are found in Cr$_3$C carbide [Figure \ref{fig2}(d)], although the curve of $\Delta$E versus magnetization is slightly altered for different strains. In contrast to the strain -- free phase transition, a second -- order phase transition occurs at an early stage of Cr substitution when x = 2 (Cr$_2$FeC).

\begin{figure}
\includegraphics[angle=-90, width=10.0cm]{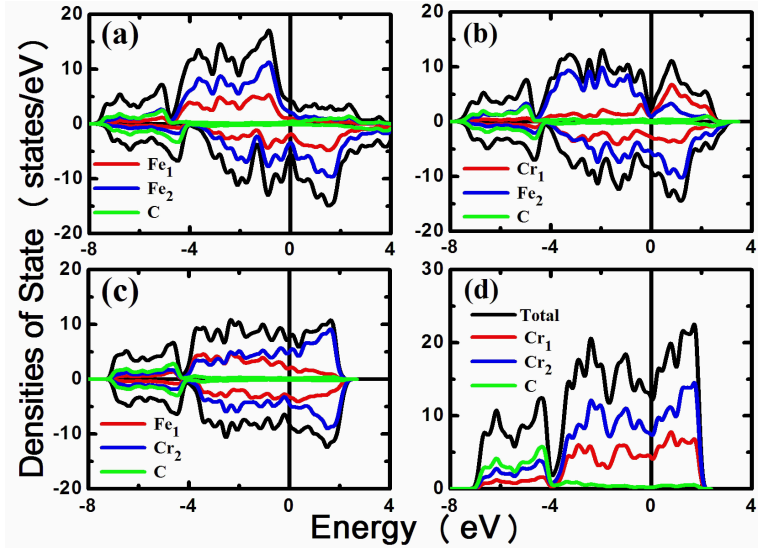}
\caption{(Color online) Total and orbital -- projected DOS of (a) Fe$_3$C, (b) Fe$_{2}$CrC, (c) Cr$_{2}$FeC and (d) Cr$_3$C. The Fermi level is set to zero energy.}
\label{fig3}
\end{figure}

The total and the orbital -- projected densities of states (DOS) for Fe$_{3-x}$Cr$_x$C compounds with x = 0, 1, 2, and 3
are plotted in Figures \ref{fig3}(a), (b), (c), and (d), respectively. While the minority spin states are presented as negatives in Figures \ref{fig3} (a) -- (c), they are positive in Figure \ref{fig3}(d) because the majority and the minority spin states of Cr$_3$C are entirely degenerate. The majority spin states of both the Fe$_1$ -- 3\emph{d} and the Fe$_2$ -- 3\emph{d} orbitals are nearly filled while the minority spin states are partially occupied [Figure \ref{fig3}(a)], analogous to the electronic features of pure Fe in bulk form. Owing to the significant spin exchange splitting, the Fe$_1$ and Fe$_2$ sites in Fe$_3$C have larger magnetic moments, but still smaller, than the bulk Fe moment (2.2 $\mu_B$), as shown in Table 1. This is due to the presence of a small portion of the majority Fe -- 3\emph{d} states in the unoccupied bands, which is induced by hybridization with the C -- 2\emph{p} states. This hybridization is also the cause of the induced magnetic moment of C atoms, as the empty C -- 2\emph{p} states in the majority spin subband shift downward towards the occupied region across the Fermi level with respect to the minority states, as seen in Figure \ref{fig3}(a). The minority spin states of Fe$_3$C almost remain in Fe$_{2}$CrC [Figure \ref{fig3}(b)], but a dense orbital -- projected DOS is found  in the majority spin state for both the Fe and the Cr sites. The coincidence of these majority spin peaks suggests that the reduced magnetic moments of the Fe and the Cr atoms accompany the strong hybridization between the Fe -- 3\emph{d} and the Cr -- 3\emph{d} orbitals. Similarly, we assign the existence of majority spin states of the Fe -- 3\emph{d} and the Cr -- 3\emph{d} orbitals above the Fermi level to the reduction in the spin exchange -- splitting at the Fe and the Cr sites of Cr$_{2}$FeC, considering that the minority spin states remain the same for all x in Fe$_{3-x}$Cr$_x$C. This spin exchange -- splitting becomes larger (smaller) when a tensile (compressive) strain is applied (not shown). Meanwhile, as shown in Figures \ref{fig3}(c) and (d), the spin subbands of the C atoms are nearly or completely degenerate, maintaining negligible net moment at x = 2 and 3 (see Table 1), respectively. Note that, the total and the orbital -- projected DOS of Cr$_3$C in Figure \ref{fig3}(d) exhibit behaviors that are similar to these minority spin states for Fe$_3$C [Figure \ref{fig3}(a)], but are degenerate in the spin subbands.

\begin{center}
{\bf IV. CONCLUSIONS}
\end{center}

In summary, using first -- principles DFT calculations, we show that the substitution of Cr atoms at Fe sites in Fe$_{3-x}$Cr$_x$C (x = 0, 1, 2, and 3) changes the magnetic ground state from a FM phase for x = 0 (Fe$_3$C) to a NM phase for x = 3 (Cr$_3$C). Two intermediate compounds when x = 1 and 2 prefer the FIM phase with an antiparallel alignment of magnetic moments between the Fe and the Cr sites. The stability of these FIM structures becomes weakened under a compressive strain along the z -- axis and is diminished for Cr$_2$FeC. This indicates that the magnetism and the magnetic phase transition in  cementite (Fe$_3$C)can be modulated by controlling both the concentration of Cr impurities and the strain effect, which may provide interesting prospects for tailoring the magnetic phase transition and the temperature in magnetocaloric and thermomagnetic applications.

\begin{center}
{\bf ACKNOWLEDGMENTS}
\end{center}

This work was supported by Incheon National University Grant in 20141336.

\newpage

\begin{table}
\caption{Optimized lattice constants, magnetic energy $\Delta$E between the FM/FIM and the NM states ($\Delta$E=E$_{\textrm{tot}}$[FM/FIM] -- E$_{\textrm{tot}}$[NM])(units of eV/cell), magnetic moments for each atomic site (units of $\mu_B$/atom), and total magnetizations of Fe$_3$C, Fe$_2$CrC, Cr$_2$FeC, and Cr$_3$C (units of $\mu_B$/f.u.).
Results from previous theoretical \cite{Lv} and experimental works \cite{Fasiska, Wood, Inoue} are shown in parentheses for comparison.}
\begin{ruledtabular}
\begin{tabular}{ccccc}
 &  Fe$_3$C &  Fe$_2$CrC &  Cr$_2$FeC &  Cr$_3$C\\
\colrule
$a$ (\AA)       &  5.05  (5.09)\cite{Fasiska, Wood} &  5.04 (5.02)\cite{Lv} &  5.13 (5.10)\cite{Lv}&  5.18 (5.12)\cite{Inoue}\\
$b$ (\AA)       &  6.73 (6.74)\cite{Fasiska, Wood} &  6.88 (6.86)\cite{Lv} &  6.57 (6.60)\cite{Lv}&  6.65 (6.80)\cite{Inoue}\\
$c$ (\AA)       &  4.49 (4.52)\cite{Fasiska, Wood} &  4.40 (4.39)\cite{Lv}&  4.51 (4.52)\cite{Lv}&  4.51 (4.58)\cite{Inoue}\\
$\Delta$E  & -2.48   & -0.80  & -0.13   & 0.0    \\
M(Fe$_1$/Cr$_1$)  &  2.09 (1.92)\cite{Lv} &  -1.25 (-1.37)\cite{Lv} &  1.15 (0.79)\cite{Lv} &  0.0 \\
M(Fe$_2$/Cr$_2$)  &  2.04 (1.99)\cite{Lv} &  1.65 (1.65)\cite{Lv} &  -0.01  (-0.03)\cite{Lv} &  0.0  \\
M(C)  &  -0.24 (-0.24)\cite{Lv} &  -0.05 (-0.03)\cite{Lv} &  -0.02 (-0.03)\cite{Lv} &  0.0 \\
M(Total) &  5.90 (5.59)\cite{Lv}  &  2.01 (1.86)\cite{Lv} &  1.07 (0.74)\cite{Lv} &  0.0  \\
\end{tabular}
\end{ruledtabular}
\label{table1}
\end{table}

\end{document}